# Quantum Interferometric Optical Lithography: Exploiting Entanglement to Beat the Diffraction Limit

Agedi N. Boto[1], Pieter Kok[2], Daniel S. Abrams[1], Samuel L. Braunstein[2], Colin P. Williams[1], and Jonathan P. Dowling[1,*]

[1]*Jet Propulsion Laboratory, California Institute of Technology*
*Mail Stop 126-347, 4800 Oak Grove Drive, Pasadena, California 91109*

[2]*Informatics, University of Wales, Bangor LL57 1UT, United Kingdom*

## Abstract

Classical optical lithography is diffraction limited to writing features of a size $\lambda/2$ or greater, where $\lambda$ is the optical wavelength. Using nonclassical photon-number states, entangled $N$ at a time, we show that it is possible to write features of minimum size $\lambda/(2N)$ in an $N$-photon absorbing substrate. This result allows one to write a factor of $N^2$ more elements on a semiconductor chip. A factor of $N = 2$ can be achieved easily with entangled photon pairs generated from optical parametric downconversion. It is shown how to write arbitrary 2D patterns by using this method.

arXiv:quant-ph/9912052 v2  4 May 2000

Optical lithography has been the primary tool of the semiconductor industry for transferring circuit images onto substrates. However, diffraction effects in the typical masking approach limits the minimal resolvable feature size to the Rayleigh diffraction limit of $\lambda/2$, where $\lambda$ is the optical wavelength. Hence, it has become necessary to use light of ever-shorter wavelengths to fabricate smaller features. Current production technology writes 180–220 nm features using KrF excimer laser light at 248 nm. New technological approaches consider light in the vacuum ultraviolet or soft $x$ ray regime in order to obtain features at 100 nm or below [1]. In all cases, the light is treated classically or, equivalently, as a stream of uncorrelated photons—an approach that leads to the Rayleigh criterion. We shall demonstrate that it is possible to overcome this limit by using entangled (correlated) photon number states, which have no classical analog.

In classical interferometric lithography [2], when two coherent plane waves of laser radiation are made to intersect at an angle of $2\theta$, as shown in Fig. (1), interference fringes form with a spacing (pitch) of $p = (\lambda/2) \sin\theta$. In the grazing incidence limit, $\theta \to \pi/2$, the minimum linear feature size that can be written is $x^{\min} = \lambda/2$. To see this, realize that the normalized exposure dose $\Delta$ at the substrate (proportional to the intensity) is given by the scaled interference pattern of two counter-propagating, grazing incidence, plane waves, $\Delta(x) = 1 + \cos(2kx) = 1 + \cos 2\varphi$. Here, $k = 2\pi/\lambda$ is the optical wavenumber, $\varphi = kx$ is the associated path-differential phase, and $x$ the lateral dimension on the substrate. (We shall assume grazing incidence $\theta \to \pi/2$ for the rest of this work.) The Rayleigh criterion states that the minimal resolvable feature size occurs at a spacing corresponding to the distance between an intensity maximum and an adjacent intensity minimum [3]. The criterion then demands $\varphi^{\min} = \pi/2$, from which we obtain $x^{\min} = \lambda/2$, as given above. This is the best resolution that can be achieved classically [2].

Recently, Yablonovitch and Vrijen (YV) have proposed utilizing "classical" two-photon exposure techniques to improve resolution [4]. The idea is that the uncorrelated (classical) two-photon absorption probability scales quadratically with the intensity [4]. Hence, the classical two-photon exposure dosage has the form, $\Delta^c_{2\gamma} = \Delta^2(x)/2 = (1+\cos 2\varphi)^2 / 2 = \frac{3}{4} + \cos 2\varphi + \frac{1}{4}\cos 4\varphi$. This function has a term $\cos 4\varphi$ that oscillates in space with twice the frequency as the single-photon function, $\Delta(x)$. If the middle term containing the more slowly oscillating $\cos 2\varphi$ could somehow be eliminated, one would be left with a pattern resolution of $x^{\min}_{2\gamma} = \lambda/4$; a factor of two improvement. Since the number of elements writeable on a surface scales inverse-quadratically with the minimum feature dimension, this is an important advance. The approach of YV is to eliminate this middle term using a classical frequency modulation scheme. In the rest of this work, we show instead how to employ entangled photon number states to improve on this result.

It has been known for some time that entangled photon pairs, such as those generated by spontaneous parametric down conversion [4], have unusual resolving characteristics. This feature allows for sub-shot-noise interferometric phase measurement in theory [6, 7] and experiment [8]. In particular, Fonseca, *et al.*, recently observed the resolution of a two-slit diffraction pattern at half the Rayleigh limit in a coincidence-counting experiment [8]. What we will now show is that this type of effect is possible not only in coincidence experiments, but also in real two-photon imaging systems. Quantum entanglement is the resource that allows sub-diffraction-limited lithography.

Consider the schematic set up for quantum interferometric lithography, illustrated in Fig. (1). We consider a two-port device with photons incident on a symmetric, lossless, beam splitter (BS) from the left in one or both ports $A$ or $B$. The photons are then reflected by a mirror pair (M) onto the substrate (S) at the right. We can model the phase differential due to path-length differences between the upper and lower branches of the interferometer as a single phase shifter (PS) placed in the upper branch, which imparts a phase shift $\varphi = kx$. The two, photon paths converge on the imaging plane at an angle of $\theta$ off the horizontal axis, as show. We identify with the two input ports $A$ and $B$, two photon annihilation operators $\hat{a}$ and $\hat{b}$, respectively [7]. These operators obey the usual photon commutation rules, $[\hat{a},\hat{a}^\dagger]=[\hat{b},\hat{b}^\dagger]=1$ and $[\hat{a},\hat{b}]=0$. We can take the output electric field operator at the image plane on the right to be proportional to the sum of two output operators $\hat{c}$ and $\hat{d}$ from the upper and lower branches of the interferometer, $C$ and $D$, respectively. Then the linear relationship between the two inputs and the two outputs can be expressed by a two-dimensional matrix equation, $\hat{\mathbf{T}}\begin{bmatrix}\hat{a}\\\hat{b}\end{bmatrix}=\begin{bmatrix}\hat{c}\\\hat{d}\end{bmatrix}$, where $\hat{\mathbf{T}}$ is the input-output matrix. For our purposes, $\hat{\mathbf{T}}$ can be thought of as being the product of matrices of the form, $\hat{\mathbf{B}}=\frac{1}{\sqrt{2}}\begin{pmatrix}-1 & i\\ i & -1\end{pmatrix}$, $\hat{\mathbf{R}}=\begin{pmatrix}-1 & 0\\ 0 & -1\end{pmatrix}$, and $\hat{\mathbf{P}}=\begin{pmatrix}e^{i\varphi} & 0\\ 0 & 1\end{pmatrix}$, which represent the unitary actions of the BS, M, and PS, respectively. Here, we have assumed a $\pi$ phase shift on each reflection off a beam splitter or a mirror and a $\pi/2$ phase shift upon transmission through the BS, as required by reciprocity [9]. We assume that the phase differential between the two paths in Fig. (1) is represented in the single parameter $\varphi$ of the PS. Hence, for Fig. (1), we have $\hat{\mathbf{T}}=\hat{\mathbf{P}}\hat{\mathbf{R}}\hat{\mathbf{B}}$. Relating input to output, we now deduce the output operators as $\hat{c}=(\hat{a}-i\hat{b})e^{i\varphi}/\sqrt{2}$ and $\hat{d}=(-i\hat{a}+\hat{b})/\sqrt{2}$. Hence, the total, scaled, electric-field annihilation operator $\hat{e}$, at S, is given by $\hat{e}=\hat{c}+\hat{d}=\frac{1}{\sqrt{2}}\left(-i+e^{i\varphi}\right)\hat{a}+\frac{1}{\sqrt{2}}\left(1-ie^{i\varphi}\right)\hat{b}$. Now, to compute the normalized, quantum, two-photon, exposure dosage $\Delta_{2\gamma}$, it is sufficient to compute the expectation values of moments of $\hat{e}$ and $\hat{e}^\dagger$, since $N$-photon absorption rates at the imaging surface will be proportional to the expectation values of $\hat{\delta}_N \equiv (\hat{e}^\dagger)^N(\hat{e})^N/N!$ the dosing operator. The quantum theory of uncorrelated two-photon absorption was first worked out in 1931 by Maria Göppert-Mayer [10]. More recently, Javanainen and Gould [11] reported the equivalent theory for entangled two-photon absorption, while Perina, Saleh, and Teich [12] developed the entangled $N$-photon theory. The expectation values of $\hat{\delta}_N$ are taken with respect to input states that enter at the left of Fig. (1). The advantage of our operator-based approach is that all of the properties of the interferometer are encoded into the operator form of $\hat{e} = \hat{c} + \hat{d}$, and one may then use different input states without having to recalculate the effects of the interferometer for each state [7].

Consider the input state $|\psi_I\rangle = |1\rangle_A |0\rangle_B$, where uncorrelated photons are incident one-at-a-time on the upper port $A$ (Fig. 1). For this state we have a deposition rate of $\Delta_{1\gamma}(\varphi) = \langle \psi_I | \hat{\delta}_1 | \psi_I \rangle = 1 - \sin 2\varphi = 1 + \cos(2\varphi + \frac{\pi}{2})$, which is the usual classical result, up to an unimportant phase shift. (Since the photons are uncorrelated, the interference pattern is the same as for $|\alpha\rangle_A |0\rangle_B$, a classical coherent-state input [7].) Notice that $\Delta_{1\gamma}(\varphi)$ has a mean value of one, which normalizes the optical power deposited per unit length to unity. The classical two-photon deposition rate is then, $\Delta^c_{2\gamma} = (1 + \cos 2\varphi)^2 / 2 = \frac{3}{4} + \cos 2\varphi + \frac{1}{4}\cos 4\varphi$, dropping the phase factor. We wish to use quantum interference to eliminate the slowly varying middle $\cos 2\varphi$ term. A simple choice of a nonclassical number-product state accomplishes this is $|\psi_{II}\rangle = |1\rangle_A |1\rangle_B$. This state is the natural output of a single-photon parametric down-conversion event [4]. The deposition rate for this state on a two-photon absorbing substrate has the form $\Delta^q_{2\gamma} = \langle \psi_{II} | \hat{\delta}_2 | \psi_{II} \rangle = 1 + \cos 4\varphi$, as desired. The term containing the slowly oscillating $\cos 2\varphi$ has been eliminated, and we are left with only the $\cos 4\varphi$ term with the resolution of $x^{\min}_{2\gamma} = \lambda / 4$. In the inset of Fig. (1), we plot the classical one-photon pattern (dashed), the classical two-photon pattern (dotted), and our entangled two-photon (plain) results. Note that our quantum pattern has narrower features and half the peak-to-peak spacing of either classical curve.

To understand the physics behind this improvement, as well as the role of entanglement, we adapt an argument used by Huelga, *et al.*, in the context of atomic-clocks [13]. (A two-port Mach-Zehnder interferometer is isomorphic to a two-pulse Ramsey atomic-clock interferometer, under the SU(2) spinor rotation group [6].) Important to note is the form of the quantum state inside the interferometer in Fig. (1), after BS but before PS, at the points $A'$ and $B'$. Interference effects upon passage through a symmetric, lossless beamsplitter, cause the product number state $|\psi_{II}\rangle = |1\rangle_A |1\rangle_B$ to become the entangled number state $|\psi_E\rangle = (|0\rangle_{A'} |2\rangle_{B'} + |2\rangle_{A'} |0\rangle_{B'}) / \sqrt{2}$. Here, the entanglement is between photon number and path; it is not possible to tell whether *both* photons took the lower path or *both* took the upper path [5]. This entangled state is sometimes called a "diphoton" or a "two-photon", and it behaves as single quantum-mechanical object, of photon-number two [14]. There are two indistinguishable paths this diphoton can take through the interferometer, and so the amplitudes corresponding to these two paths will interfere. However, the diphoton will pass through the phase shifter only on the upper path, and this probability amplitude will acquire *twice* the phase shift as with a single photon process. Therefore, it is easy to see that the entangled state becomes, $|\psi_E(\varphi)\rangle = (|0\rangle_C|2\rangle_D + e^{2i\varphi}|2\rangle_C|0\rangle_D)/\sqrt{2}$. This is the origin of the doubling of the resolution in the deposition rate $\Delta_{2\gamma}^q$, given above.

We can compute this same rate $\Delta_{2\gamma}^q$ by a slightly more general argument. Suppose we ignore the process by which the two-photon entangled state is created in the interferometer, for the moment, and simply assume that the state $|\psi_E(\varphi)\rangle$ has the maximally entangled form, given above, just to the right of PS. It now is not necessary to relate the output to the input operators. Instead, we can compute the deposition expectation function directly as $\langle\psi_E(\varphi)|\hat{\delta}_2|\psi_E(\varphi)\rangle$, using $\hat{e} = \hat{c} + \hat{d}$, which gives the same result $\Delta_{2\gamma}^q$ as before. It is now a simple matter to show that if an entangled number state of the form $|\psi_E(N)\rangle = (|0\rangle_{A'}|N\rangle_{B'} + |N\rangle_{A'}|0\rangle_{B'})/\sqrt{2}$ is prepared inside the interferometer at the points $A'$ and $B'$ (Fig. 1), then the phase from the path differential accumulates $N$ times as fast, producing $|\psi_E(N,\varphi)\rangle = (|0\rangle_C|N\rangle_D + e^{iN\varphi}|N\rangle_C|0\rangle_D)/\sqrt{2}$ at the output ports [13]. If the substrate is $N$-photon absorbing, then the deposition function may be directly computed as,

$$\Delta_N(\varphi) \equiv \langle\psi_E(N,\varphi)|\hat{\delta}_N|\psi_E(N,\varphi)\rangle = 1 + \cos 2N\varphi \ . \qquad (1)$$

All the slowly oscillating terms have been eliminated, leaving only the term $\cos 2N\varphi$ that has a resolution of $\lambda/(2N)$. This is a remarkable factor of $N$ below the classical, Rayleigh limit of $\lambda/2$.

Of course, the natural question arises: how can one produce maximally entangled photon number states $|\psi_E(N)\rangle$ of this form? For the case of $N = 2$, the answer is a parametric down-conversion event followed by a symmetric beamsplitter. For higher $N$, one approach is to use a material with a $\chi^{(N)}$ nonlinearity, in which a single photon is downconverted into $N$ entangled daughter photons [12]. Another approach is to utilize a cascaded arrangement of $N–1$ crystals with $\chi^{(2)}$ nonlinearities [12, 15]. It is interesting to note that the natural output of a series of parametric downconversion events, or that of an optical parametric oscillator (OPO), has the form $|N/2\rangle_A|N/2\rangle_B$ of a Hilbert space product [6,7]. In the case $N = 2$, *only* a simple linear beam splitter need be used to generated the maximally entangled form of $|\psi_E(N)\rangle$. More generally, for $N > 2$, one can construct such a state by operating on $|N/2\rangle_A|N/2\rangle_B$ with a series of $N$, nonlinear, photon-photon, controlled-NOT gates [13, 16].

It is clear that the procedure described above is sufficient for etching a series of parallel lines with a factor of $N$ reduction in the pitch. For more general-purpose lithography, one would like to write arbitrary patterns in 2D. It is a simple matter to extend our technique to do this. Our approach makes use of the most general entangled state that can be generated by the above-mentioned techniques for fixed $N$, namely, $|\psi_{N,P}(\varphi)\rangle = \left(e^{iP\varphi}|N-P\rangle_A|P\rangle_B + e^{i(N-P)\varphi}|P\rangle_A|N-P\rangle\right)$. With phase shifters, we can combine two such states interferometrically as $|\Psi_{N,P,P'}\rangle = \alpha_{N,P}|\psi_{N,P}\rangle + \beta_{N,P'}|\psi_{N,P'}\rangle$, with complex coefficients α and β such that $|\alpha|^2 + |\beta|^2 = 1$. This set of states can now be used as a basis to expand the target intensity pattern $p(\varphi)$ in terms of the deposition components $\Delta_N^{P,P'} \equiv \langle\Psi_{N,P,P'}|\hat{\delta}_N|\Psi_{N,P,P'}\rangle$. The task now is to calculate an appropriate set of these coefficients in order to fit a target function $p(\varphi)$. This can be done using a genetic algorithm optimization scheme [17]. For an example of this approach, let us approximate a 1D square well "trench", given by the sample target function $p(\varphi) \in \equiv 1 - \in \Theta(\varphi - \pi/2) + \Theta(\varphi - 3\pi/2)$ where $\Theta$ is the unit step function. The target function $p(\varphi)$ is depicted as a dark solid line in Fig. (2). The dotted line shows the best possible one-photon classical interferometric approximation. The solid curve is our quantum interferometric approximation, expanded with $P \in \{1,...,5\}$ while $N = 10$, and optimized with a genetic algorithm. Note that our entangled pattern (solid curve) is a much better approximation to the target trench function (dark solid line) than is the best classical approximation (dashed curve). It is a simple matter to extend this technique to 2D by adding orthogonal photon modes, as we shall show in a forthcoming work [17]. This method for producing arbitrary patterns in 2D is analogous to techniques used in commercial lithography—the Fourier components of the desired image are manipulated using phase masks, to insure that the actual image is as close to the target image as possible [18].

We make one final important note on the entangled $N$-photon absorption process. Classically, the uncorrelated $N$-photon absorption probability scales like $I^N$, where $I$ is the normalized classical intensity. This result is a consequence of the fact that the photons arrive independently of one another. Hence, the probability that the first photon arrives in an elemental absorption volume in space-time is proportional to $I$, and the probability a second photon will also happen to be in the same volume is also proportional to $I$, and so on, giving the $I^N$ law. For this reason, $N$-photon absorption lithography with uncorrelated "classical" light is infeasible for high $N$, since extremely high flux densities are required [12]. This is not the case for $N$-photon absorption with entangled photons. For two-photon absorption with entangled photon pairs, the absorption cross section scales as $I$, and not the $I^2$ that would be expected classically [11]. In general, this scaling law holds for arbitrary $N$. In other words, the $N$-photon absorption cross-section, with $N$ entangled photons, scales like $I$ and not $I^N$ expected classically [12]. This result can be seen by the following heuristic argument: Recall that the photons are correlated in space and time, as well as number. Hence, if the optical system is aligned properly, the probability of the first photon arriving in a small absorptive volume of spacetime is proportional to $I$. However, the remaining $N-1$ photons are constrained to arrive at the same place at the same time, and so each of their arrival probabilities is a constant, independent of $I$. Hence, although classical $N$-photon lithography requires unrealistically high optical powers, entangled $N$- photon lithography requires only the same levels of power as the classical one-photon device.

In conclusion, we have discussed the problem of entangled $N$-photon absorption, as applied to interferometric optical lithography. We conclude that highly entangled photon states can be used to write features in an $N$-photon absorbing resist which have a resolution of $\lambda/(2N)$. This result is a factor of $N$ below the classical Rayleigh diffraction limit. Such states can easily be made for $N = 2$ using optical parametric downconversion, and there are several possible approaches for implementing the scheme for higher $N$. We also indicate how the technique may by extended to write arbitrary 2D patterns. Entanglement turns out to be a useful resource, which can be employed in a technology such as lithography to overcome seemingly unbreakable constraints such as the diffraction limit.

## ACKNOWLEDGEMENTS

We would like to acknowledge useful discussions with R. Y. Chiao, J. H. Eberly, J. D. Franson, H. J. Kimble, P. W. Kwiat, H. Mabuchi, M. O. Scully, A. Scherer, J. E. Sipe, R. B. Vrijen, and E. Yablonovitch. This work was carried out under a contract with the National Aeronautics and Space Administration. Also, SLB and PK acknowledge partial support from the UK Engineering and Physical Sciences Research Council.

**FIGURE CAPTIONS**

Fig. 1    Interferometric lithography set up utilizing photons entering ports A and B. The photons strike the symmetric, lossless, beamsplitter (BS) and then reflect off the mirrors (M). The photon amplitude in the upper path accumulates a phase shift $\varphi$ at the phase shifter (PS), before the two branches interfere on the substrate. Inset: The deposition rate $\Delta$ as a function of the phase shift $\varphi$ for uncorrelated single-photon absorption (dashed), uncorrelated two-photon absorption (dotted), and entangled two-photon absorption (solid). Note that the classical two-photon curve has narrower features than the one-photon, but that the entangled two-photon has even narrower features still, and also shows the critical halving of the peak separation.

Fig. 2    Here we show a target "trench" function (dark solid line), the best classical interferometric approximation (dashed), and our quantum-entangled interferometric approximation with ten entangled photons.

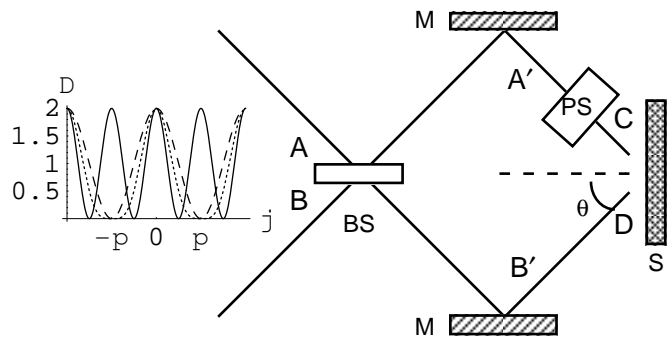

Fig. 1

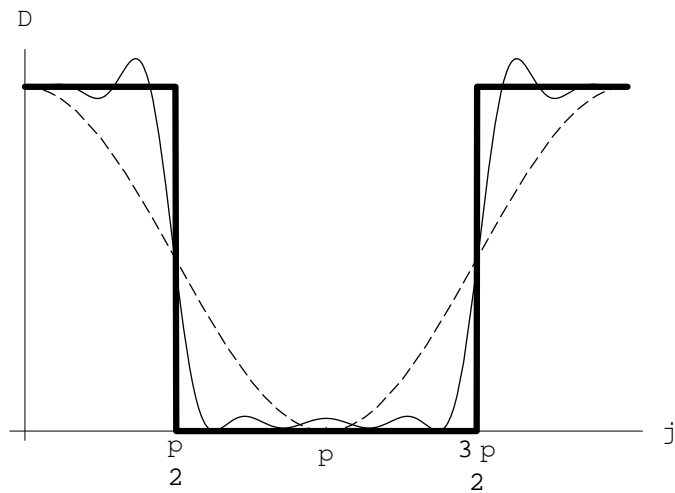

Fig. 2